\documentclass[journal=nalefd,manuscript=article]{achemso}

\title{High-speed discrimination and sorting of sub-micron particles using a microfluidic device}

\author{Sukumar Rajauria}
\altaffiliation{Present address: HGST Western Digital, San Jose Research Center, San Jose, California 95138 USA}
\affiliation{Department of Physics, University of California, Santa Barbara, California 93106 USA}
\author{Christopher Axline}
\altaffiliation{Present address: Department of Applied Physics, Yale University, New Haven, Connecticut 06520 USA}
\affiliation{Department of Physics, University of California, Santa Barbara, California 93106 USA}
\author{Claudia Gottstein}
\affiliation{Neuroscience Research Institute, University of California, Santa Barbara, California 93106 USA}
\author{Andrew N. Cleland}
\email{anc@uchicago.edu}
\altaffiliation{Present address: Institute for Molecular Engineering, University of Chicago, Chicago Illinois 60637 USA}
\affiliation{Department of Physics and California Nanosystems Institute, University of California, Santa Barbara, California 93106 USA}


\date{\today}

\begin{document}

\begin{tocentry}

\includegraphics[width=6cm]{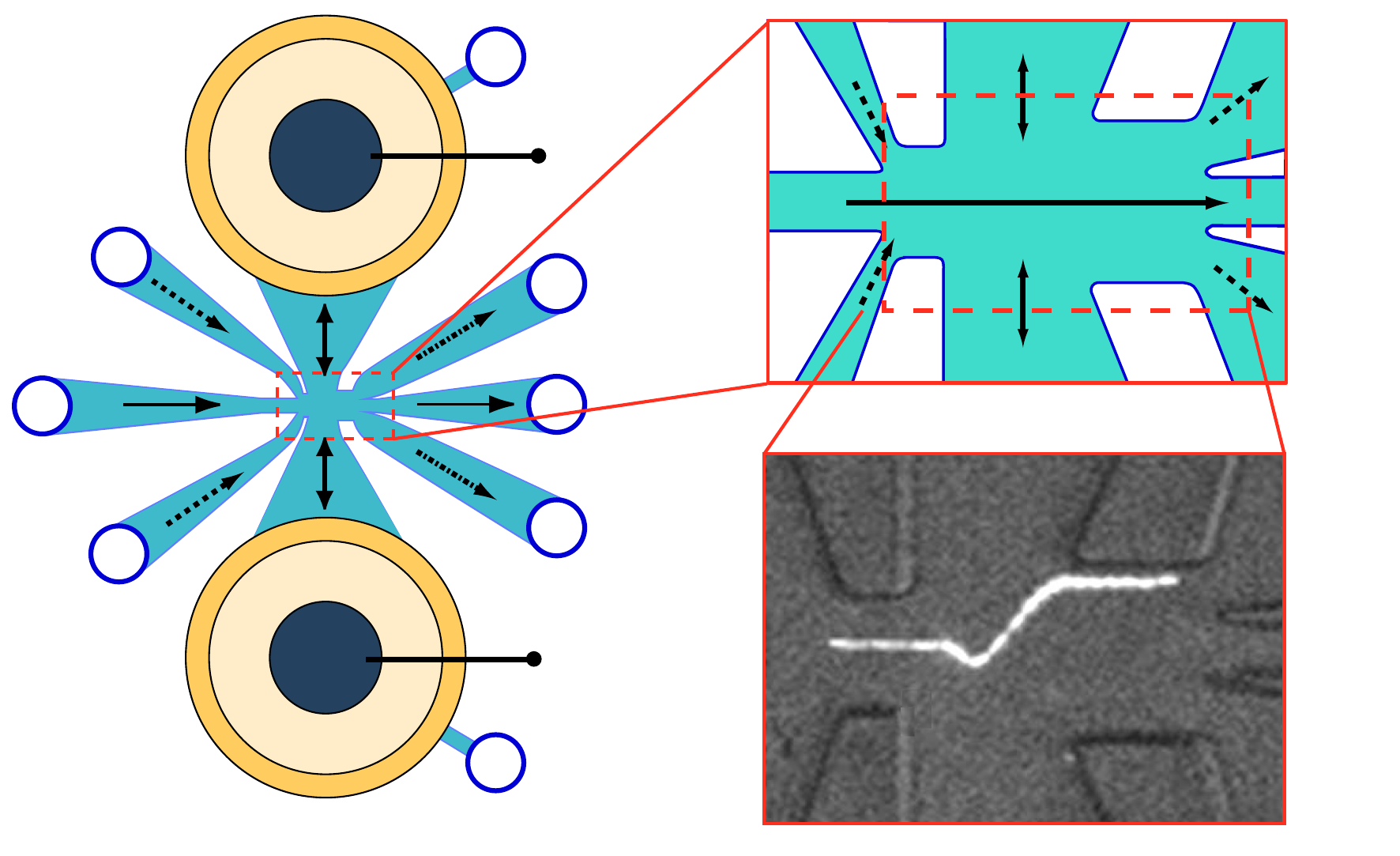}

\end{tocentry}

\clearpage
\begin{abstract}
\noindent
The size- and fluorescence-based sorting of micro- and nano-scale particles suspended in fluid presents a significant and important challenge for both sample analysis and for manufacturing of nanoparticle-based products. Here we demonstrate a disposable microfluidic particle sorter that enables high-throughput, on-demand counting and binary sorting of sub-micron particles and cells, using either fluorescence or an electrically-based determination of particle size. Size-based sorting uses a resistive pulse sensor integrated on-chip, while fluorescence-based discrimination is achieved using on-the-fly optical image capture and analysis. Following detection and analysis, the individual particles are deflected using a pair of piezoelectric actuators, directing the particles into one of two desired output channels; the main flow goes into a third waste channel. The integrated system can achieve sorting fidelities of better than 98\%, and the mechanism can successfully count and actuate, on demand, more than 60,000 particles/min.
\end{abstract}

\maketitle


\clearpage

Synthetic and naturally-occurring nanoparticles are playing an increasingly significant role in both research and industry. In medicine, nanoparticles are increasingly being developed for therapeutics as well as implicated in disease \cite{AlivisatosNatureBio04, MichaletScience05, HuangNanomedicine07, TirrellNature04, RosenbergAgg06}; in manufacturing, nanoparticles are being developed for data storage \cite{SunScience00, HyeonCC03, LuAC07}, photovoltaics \cite{SeowNanotechnology09, YellaScience11}, as well as food additives, cosmetics, and paint \cite{WeirFoodEST2012}. Many applications rely on nanoparticles of a given size; however the methods used to generate nanoparticles tend to produce a distribution of diameters. The ability to rapidly count and sort nanoparticles based on size therefore presents a compelling approach to analyzing and narrowing these distributions. More generally, a technology that can actively separate nano- and micro-particles based on fluorescence, magnetic response, or other physical attributes in addition to size would be a highly useful tool for both nanoparticle synthesis and analysis.

A variety of methods for sorting micro- and nanoparticles have been demonstrated previously. To date, these implementations have been limited in speed and fidelity \cite{HuangScience04, BaretLabChip10, HanScience00, StavisLabChip10, FuNatureBio99, MiltenyiCytometry90, LeeAPL01}. Here we describe an alternative mechanism for the high-speed, high-fidelity binary sorting of sub-micron particles, using particle size determined by resistive pulse sensing (RPS) \cite{DeBloiRSI70} to detect and discriminate.  Following discrimination, particles are actuated using an on-chip piezoelectric push-pull microsorter that deflects individual particles into the desired output channel \cite{ChenBiomedMicro09, ZhouMicromachines11}. We also demonstrate that the microactuator can be used to count and sort particles based on fluorescence, thereby providing a means for the gentle and reliable binary sorting of e.g. populations of biological cells.

We first describe the resistive pulse sensing and sorting approach, in which size analysis and on-demand actuation are integrated in a single device. We demonstrate high-speed sorting of a binary mixture of sub-micron diameter polystyrene beads, and also use this device to test the maximum rate at which we can accurately actuate particles. Next, we describe how we construct a fluorescence-based sorter, using the optical signals from fluorescent microparticles as a sorting trigger for the same microactuator design. We use this approach to demonstrate counting and binary sorting of a mixture of fluorescently stained and un-stained mammalian cells. This provides a gentle and reliable alternative to detection and sorting compared to e.g. conventional flow cytometry.

\section*{Results}

\subsection*{Electrically-based size detection and discrimination}

\begin{figure*}
\includegraphics[width=\textwidth]{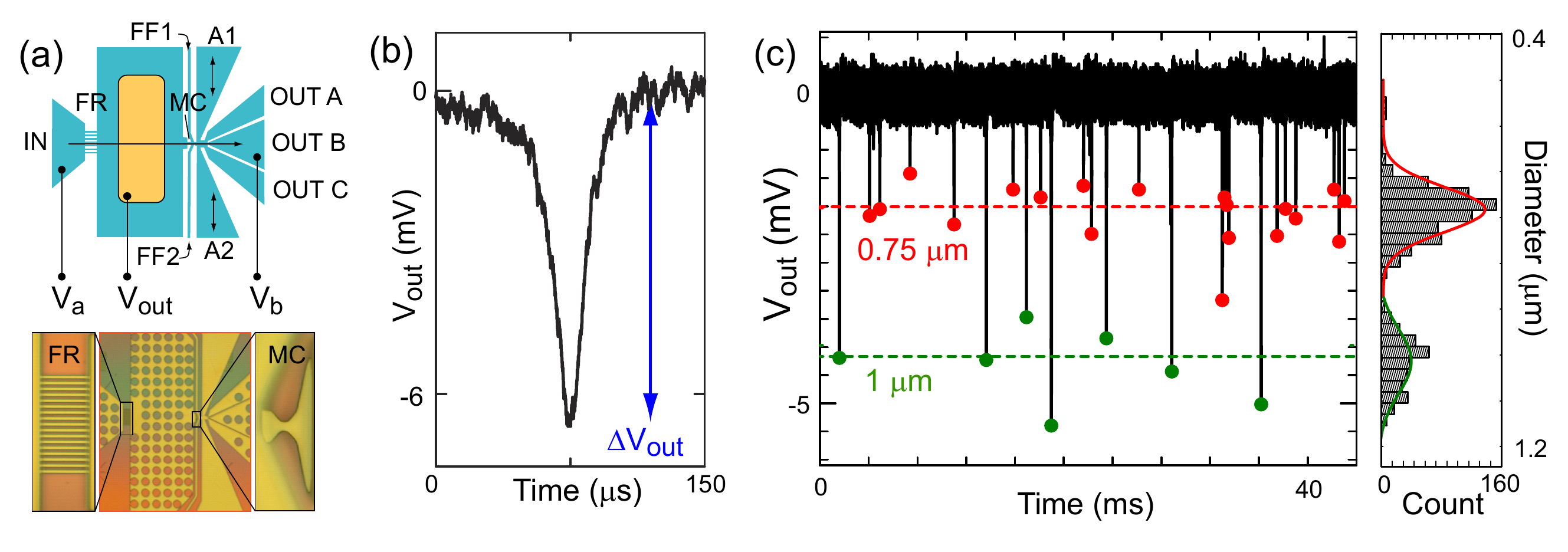}
\caption{\textbf{Electrical detection of particles}: (a) Electrical particle size analyzer. Upper drawing shows layout, with input flow (IN) passing through a fluidic resistor (FR) and a microconstriction (MC), in between which a sensor electrode senses the output voltage $V_{\rm out}$ due to the bias $V_a$ and $V_b$. Following the microconstriction, particles are flow-focused (FF1 and FF2) then actuated (A1 and A2) into one of the three output channels. Image below shows the microfluidic device with magnified views of the fluid resistor and microconstriction.  (b) Output voltage $V_{\rm out}$ as a function of time when a single 1 $\mu$m diameter latex bead passes through the microconstriction, with good time resolution and signal-to-noise ratio. The peak voltage change is proportional to the applied potential difference $V_b-V_a$ across the microconstriction as well as to the volume of the particle. (c) Left: Accumulated voltage pulses from a mixture of 1 $\mu$m and 0.75 $\mu$m diameter beads. Right: Histogram of Gaussian-fit voltage amplitudes from data similar to that shown on left.  \label{fig:1}}
\end{figure*}

Resistive pulse sensing has been demonstrated for the analysis of particle size down to particles a few tens of nanometers in diameter \cite{SalehNanoLetters03, LiNatureMat03, DekkerNatureNano07}, and has been implemented with quite large measurement bandwidth, enabling particle detection at rates approaching $10^5$ particles/s \cite{WoodAPL05, WoodRSI07, FraikinNatureNano11}. Our high-speed nanoparticle analyzer \cite{FraikinNatureNano11} is shown in Fig. \ref{fig:1}(a), comprising a molded microconstriction (MC), which when filled with saline has electrical resistance $R_b$, a molded microfluidic bias resistor (FR), with saline electrical resistance $R_a$, and a metal electrode that capacitively senses the saline electrostatic voltage $V_{\rm out}$ between the microconstriction and bias resistor. When the device is biased by an externally-applied voltage difference $V_b-V_a$, and a particle passes through the microconstriction, the microconstriction resistance will increase by an amount $\Delta R_b$ that is proportional to $R_b$ and to the particle volume \cite{DeBloiRSI70}. This changes the electrostatic potential of the saline above the metal sense electrode by an amount
\begin{equation}\label{eq.1}
    \Delta V = -\frac{R_a R_b}{(R_a+R_b)^2} \, \frac{\Delta R_b}{R_b} \, \left (V_{b}-V_{a}\right ).
\end{equation}
In an optimally-designed circuit the MC and FR resistances are balanced, $R_b = R_a$, and the first term in Eq. (\ref{eq.1}) is equal to $1/4$. In the limit where the amplifier circuit connected to the electrode has infinite input impedance, the voltage change at the amplifier input is given by Eq. (\ref{eq.1}), yielding a large bandwidth voltage signal proportional to the particle volume.

Figure \ref{fig:1}(b) demonstrates the electrical detection of a single 1 $\mu$m diameter polystyrene particle. It shows both the high sensitivity and the rapid electrical response of the device, corresponding to the short transient ($\sim 100~\mu$s) generated by the particle passage through the microconstriction. The amplitude of the voltage signal can be used to discriminate particle size, as in Fig. \ref{fig:1}(c), showing signals generated by a mixture of 0.75 $\mu$m and 1 $\mu$m diameter polystyrene beads. Two distinct signal amplitudes are visible in the time trace, which are accumulated to generate the histogram also shown in Fig. \ref{fig:1}(c).

\subsection*{Particle actuation}
\begin{figure*}[htp]
\includegraphics[width=\textwidth]{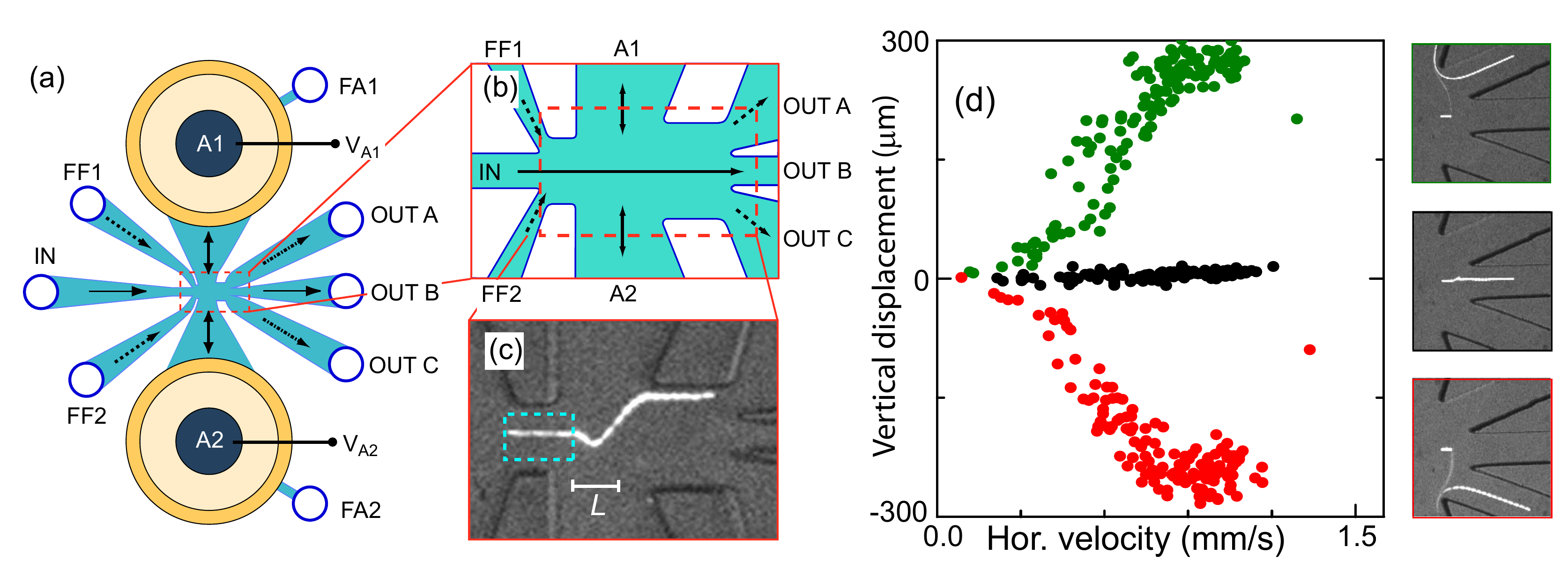}
\caption {\textbf{Push-pull actuator:} (a) Overview: Primary flow is from input (IN) to waste port (OUT B). Focusing flow (FF1 and FF2, dashed arrows) confines the particle lateral distribution in main flow. A pair of disc transducers (A1 and A2) generate fluid flow to or from the actuators (vertical arrows); a volume below each actuator (not shown) is filled using FA1 and FA2. The resulting transverse drag flow displaces particles towards output ports OUT A and OUT C (dashed-dot arrows). (b) Expanded view of sorting region in (a), roughly the microscope field of view. (c) Long exposure showing a single fluorescent polystyrene bead (0.5 $\mu$m diameter) during an actuation directing the particle to OUT A. Scale bar $L = 10 ~\mu$m. Layout for fluorescence-based sorting is similar, but with $10\times$ scale, so $L = 100 ~\mu$m; blue dashed box indicates region of interest (ROI) for fluorescence signal. (d) Individual actuator tune-up: Bead displacement when actuating just upper (red) or lower (green) actuator in push mode, with voltage adjusted to give equal bead displacement from each actuator. Driving both actuators (black) shows good balance, with no net bead displacement. Images to right show time tracks for a particle corresponding to data in green (upwards displacement, top image), black (no net displacement, middle image) and red (downward displacement, bottom image). A similar tune-up was done for the pull mode actuation. \label{fig:2}}
\end{figure*}

We actively displace particles using a pair of metal bimorph disc actuators, made from a lead zirconate-titanate piezoelectric material and purchased commercially. The actuators bend in a concave-up or concave-down fashion, determined by the sign of an externally applied voltage. These actuators are placed symmetrically on either side of the sorting region of the microfluidic device, each above a small volume of fluid connected by a narrow channel to the main channel through which particles pass (see Fig. \ref{fig:2}(a)). The actuators are used in a tandem push-pull arrangement, one driven by a voltage pulse with polarity chosen so that it pushes the underlying fluid, and the other driven by the opposite polarity voltage pulse, so that it pulls on the fluid underneath, motion that generates fluid flow with minimal static pressure change. When a particle passes through the sorting region of the device, these opposite polarity voltage pulses ($V_{A1}$, $V_{A2}$) are generated and applied to the actuators. The resulting pulse of fluid flow, transverse to the main channel flow in which the particle is entrained, displaces the particle towards one of the output channels.

The microfluidic device shown in Fig. \ref{fig:2} was fabricated by molding polydimethylsiloxane (PDMS) using a lithographically-defined mold made from photo-definable epoxy. The actuators were embedded in the PDMS structure, with a very thin PDMS layer separating the actuator from the fluid, thus yielding strong and efficient mechanical coupling to the fluid. We mounted the completed device on the stage of an inverted fluorescence microscope, with the active area of the device focussed through a dry objective onto the plane of a CCD camera. For the initial experiments we used fluorescent polystyrene beads, 0.5 $\mu$m in diameter. The microfluidic channels were completely filled with either filtered saline or deionized water, and fluid flow was established by regulating the air pressure at each of the device ports. We then introduced the beads into the input port on the device, and used parallel flow in the two flow-focussing ports on either side of the input port to hydrodynamically focus the flow, yielding bead velocities of $\sim$1 mm/s. The beads were sufficiently diluted prior to injection that only one bead passed through the active area of the device at a time.

To deflect a bead, two opposite-polarity electrical pulses with amplitude 1-5 V and duration 10 ms were applied to the transducers, generating the desired push-pull drag flow in the fluid channel connecting the two transducers. Figure \ref{fig:2}(c) displays the resulting response of an individual bead to this kind of signal, with the bead directed to output port A. Reversing the polarity of the signals would instead deflect the particle to output port C. In the absence of an actuation signal, particles would flow into the waste port B.

\section*{Size-based sorting of particles}

\begin{figure*}[htp]
\includegraphics[width=0.4\textwidth]{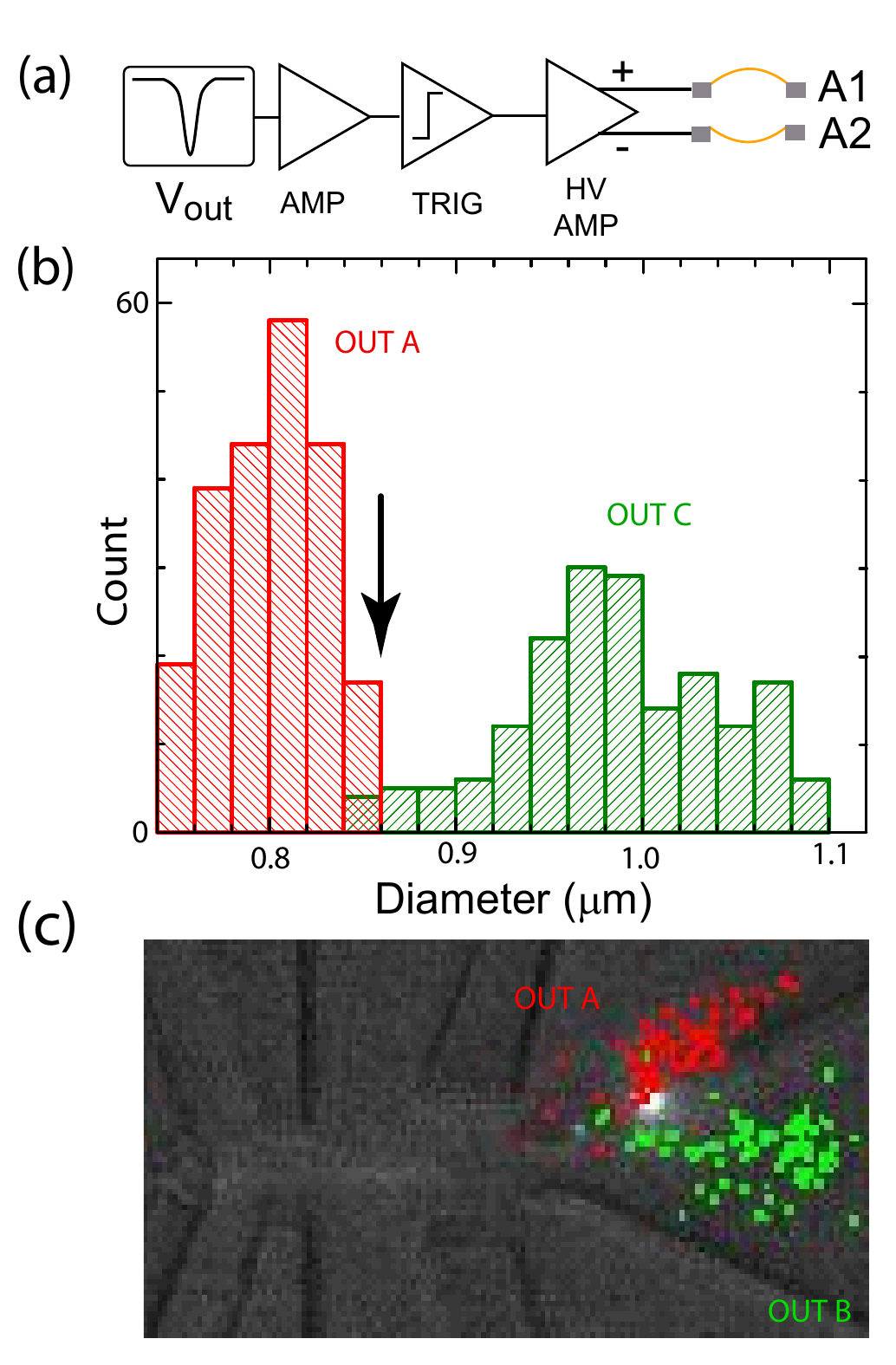}
\caption {\textbf{Electrical sorting of microparticles:} (a) Schematic for size-based microsorter, which performs a binary sort based on particle size. The electrical particle detection signal $V_{\rm out}$ is amplified (AMP) then used to generate a delayed binary signal (TRIG) that passes through a high-voltage amplifier (HV AMP) which drives the actuators (A1 and A2). (b)  Histogram of effective particle diameters measured using the particle analyzer, for a mixture of polystyrene beads 0.75 and 1 $\mu$m in diameter. The 1 $\mu$m peak is used to calibrate the horizontal axis. The histogram is color-coded according to the sorter output, where red and green corresponds to particles sent to outputs ports A and C, respectively. The sorter was operated using a discriminator set to the voltage corresponding to a particle diameter of 0.86 $\mu$m (black arrow). (c) High speed actuation. Red and green dots show the final positions of particles with (port A) and without (port B) actuation. For the actuated particles the pulse width of the actuation signal was 500 $\mu$s, which corresponds to a sorting speed of $\sim$ 60,000 particles/min, with a fidelity of around 98\%. \label{fig:3}}
\end{figure*}

We combined the electrical size analyzer and the micro-actuator to demonstrate all-electronic size-based sorting. We established pressure-driven flow of the analyte through the main channel of the device, using flow-focussing to maintain tight lateral spacing of the particles. Particles passed one at a time through the size analyzer, resulting in the generation of an output pulse whose amplitude, proportional to the particle volume, was used to discriminate and drive the piezoelectric actuators. The microactuator flow pulse then directed the particle to the desired output port. The electrical schematic is shown in Fig. \ref{fig:3}(a); the electrical signal from the particle analyzer was discriminated and a trigger generated, after a programmable time delay, that was then amplified and used to drive the actuators.

In Fig. \ref{fig:3}(b) we show the results of a size-based sorting experiment, using a mixture of polystyrene beads with 0.75 $\mu$m and 1 $\mu$m diameters. The figure displays the input size histogram, color-coded for where the input particles were directed (red for output A, green for output C). We mark also the size (voltage) at which the binary sort discriminator and trigger was set.  A single pass through the device produces highly enriched outputs, with monodisperse populations achieving (in this experiment) 100\% sorting fidelity at each output port, defined as the fraction of beads appearing at each output port that were intentionally directed there. The fidelity of the sorter is ultimately limited by variations in the particle speed and by insufficient drag force on some particles, both of which arise for particles near the bottom and top of the channel, where the Poiseuille flow velocity and the micro-actuator induced drag both fall to zero.

Very high speed sorting can be achieved using this microactuator design. Figure \ref{fig:3}(c) shows the high-speed binary sorting of 1 $\mu m$ diameter polystyrene particles, where we significantly increased the flow rate of particles over the previous experiments. The voltage pulse from the particle size analyzer was directly amplified to generate an actuation voltage that sent the first $N$ particles to output A. The actuation voltage was then shut off, and the next $N$ particles were directed to output B. Red and green dots show the final positions of particles in the output region. An output fidelity of around 98$\%$ was achieved, as shown in Table \ref{table:1}. For the actuated particles, the actuation signal pulse width was about 500 $\mu$s, which corresponds to a maximum sorting speed of about 60,000 particles/min, assuming a 50\% duty cycle. The sorting rate is limited in this case by the compliance of the actuators; stiffer actuators would be required to achieve sorting speeds significantly higher than this.

\begin{table}[ht]
\caption{Size based sorting: Results of sorting polydisperse particles with 0.75 and 1 $\mu$m diameters, showing the number of beads in each output port as well as the sorting fidelity, defined in the text. \label{table:1}}
\begin{tabular}{ccccc}
Output port & Target size & Output & Output & Fidelity \\
  & ($\mu$m) & (1 $\mu$m) & (0.75 $\mu$m) &(\%) \\
\hline
A & 1 & 180 & 0 & 100\\
B & - & 0 & 14 & - \\
C & 0.75 & 0 & 207 & 100 \\
\end{tabular}
\end{table}

\section*{Fluorescence-based particle and cell sorting}
In addition to the size-based sorting, we also used the micro-actuator in conjunction with microscope-based fluorescent image analysis to construct a fluorescence-based sorter. This was then applied to the sorting of different size fluorescent beads, as well as the sorting of stained from unstained biological cells. The implementation and its tune-up, which required scaled-up microfluidic design to accommodate larger diameter beads and cells, are described in the Methods section. We were able to bring the fidelity of this sorter, defined as the fraction of fluorescent beads that went to the desired output port, up to a value of 94\%.

To demonstrate the ability of this device to sort biological cell populations, we used murine macrophages, a model system for stem cells, stained with a Hoechst dye. In a stem cell sorting experiment, this cell-permeable dye is rapidly transported out of stem cells, allowing discrimination of stem cells from non-stem cells through the resulting side population in e.g. flow cytometry-based assays. In our model system, where we used a single type of cells, a differentiated population was achieved by spiking the stained cells into a non-stained cell population at a ratio of 3:1. This mixture was then passed through the fluorescence-based sorter.

\begin{figure}[htp]
\includegraphics[width=0.6\textwidth]{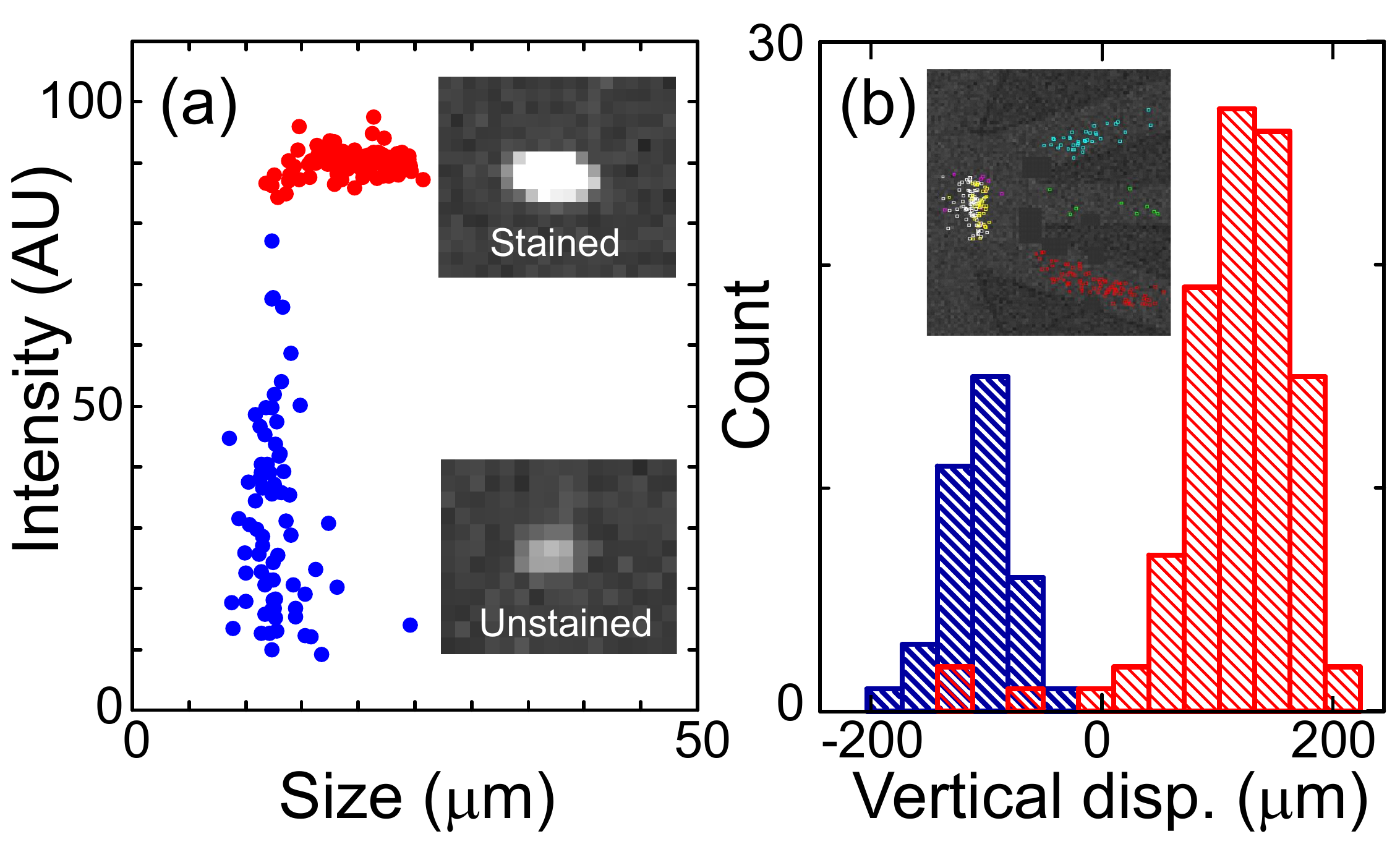}
\caption{\textbf{Fluorescence-activated cell sorting:} (a) Discrimination between Hoechst-bright (red) and Hoechst-negative (blue) mammalian cells, where each cell is evaluated for its size (width of Gaussian fit to intensity image) and for its integrated fluorescent intensity. (b) Histogram of sorting results, as a function of final vertical displacement of cells following the actuation pulse. Inset shows positions of cells before (yellow, white, magenta) and after (red, aqua, green) the sorting process. Yellow and red dots are for Hoechst-bright cells, white and aqua are for Hoechst-negative cells, and magenta and green were cells not actuated correctly.  \label{fig:4}}
\end{figure}

Figure \ref{fig:4}(a) shows the discrimination we can achieve between Hoechst-bright and Hoechst-negative cells. The Hoechst-bright cells had significantly higher intensity at the emission wavelength of 461 nm than the Hoechst-negative cells, even accounting for cell size; cell size was analyzed by evaluating the width of the Gaussian intensity profile for each cell. The intensity difference between the Hoechst-bright and Hoechst-negative cells when passed through the ROI was used to discriminate between the two cell types, and trigger the transducers accordingly. The inset to Fig. \ref{fig:4}(b) shows the initial positions of cells inside the ROI, as well as each cell's final position in the output port, where we directed the Hoechst-negative cells to port A, Hoechst-bright cells to port C, and unidentified cells to port B. A single pass sorting produces highly purified and enriched output solutions, as displayed in Table \ref{table:2}. The two output ports (A and C) received 94.5\% of the total cell population, with 6.5\% of the cells passing to the waste port B. The sorting fidelity for the cells, defined as the fraction of cells at an output port that were intentionally directed there, was 95\% at output port A, while the fidelity at output port C was 100\%.

\begin{table}[ht]
\caption{Fluorescence-activated cell sorting of Hoechst-bright versus Hoechst-negative J774 macrophage. Table shows the sorting results in numbers of cells; the last column shows the sorting fidelity, defined in the text.\label{table:2}}
\begin{tabular}{ccccc}
Output & Target& Negative cells & Bright cells & Fidelity \\
 port & cell type &  (number) & (number) & \% \\
\hline
A & Negative & 110 & 6 & 95 \\
B & Waste & 4 & 12 & - \\
C & Bright & 0 & 106 & 100 \\
\end{tabular}
\end{table}

This experiment demonstrates a powerful and potentially useful application for this device, enabling the on-demand sorting of cells based on functional parameters or based on surface or internal marker expression, following which the sorted cells can be used for further study. In the fluorescence mode, this device achieves a sorting throughput of $\sim$6,000 cells/min with a fidelity of better than 95\%, which is comparable to the results reported in the literature \cite{BaretLabChip10, ChenBiomedMicro09}. This approach promises an alternative and significantly more flexible approach to cell sorting than is possible using conventional flow cytometry.

\section*{Conclusions}

In conclusion, we have developed a fluorescence-based as well as a size-based high-speed microsorter, both implemented in a disposable microfluidic format. The two versions of the device enable discrimination of individual sub-micron particles and cells with good signal-to-noise ratio, which control an integrated push-pull micro-actuator, yielding a sorter that can be operated at high speed with high fidelity. The sorting scheme is label-free, and we have demonstrated the ability to perform binary sorts of both synthetic polystyrene beads as well as biological cells.

This concept can be extended to sort multi-component nanoparticle solutions. A multichannel output design can easily be configured to allow more complex configurations, combining both electrical and optical detection techniques as well as doing higher-order sorting, with different binary sort elements cascaded to allow multiplexed operation. A cascaded arrangement could also be used to correct sorting errors from previous sorting elements. It could also be possible to equip the CCD camera with a Bayer mask or use a triple-CCD device, in order to obtain optical images of the cells, similar to imaging flow cytometry, but with added sorting capability. The low cost, scalable fabrication, and simple on-chip detection and sorting ability, make this device potentially useful in a wide range of applications.

\clearpage

\section*{Methods}

\subsection*{Device fabrication}
\begin{figure*}
\centering
\includegraphics[width=0.8 \textwidth]{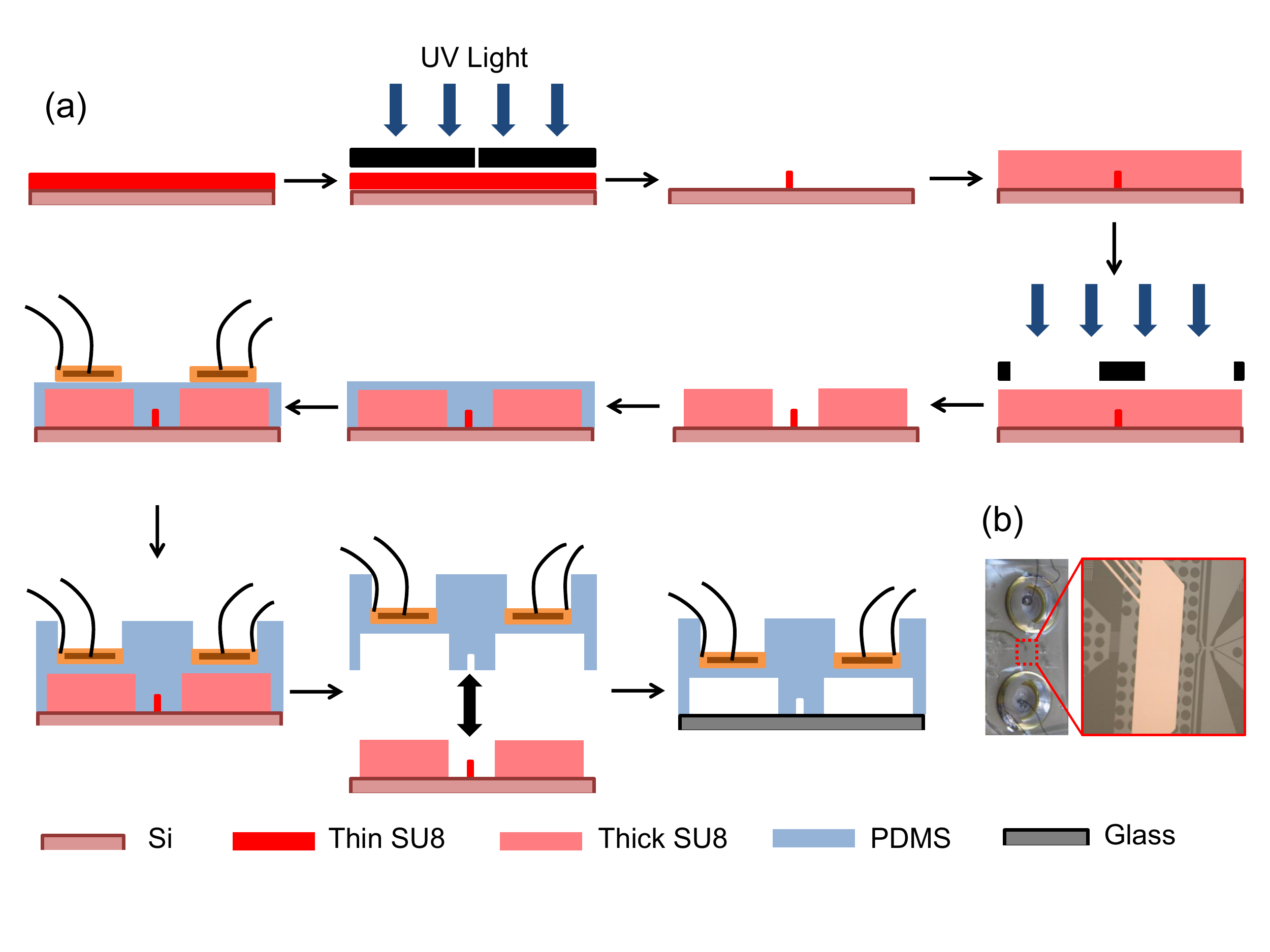}
\caption{\textbf{Fabrication process:} (a) Fabrication process flow for fabricating the PDMS-based devices. The first six steps show how the micromold was patterned, using the photodefinable epoxy SU-8 (MicroChem Corp.). The first three steps are to define the thinner (2 $\mu$m thick) fluidic resistor and microconstriction. The next three steps were used to define the thicker (15 $\mu$m) main features of the structure, including the fluidic channels and the volumes for the actuator elements. We then used to define the actual PDMS device, in two different stages. First (steps 7 and 8), a very thin PDMS layer was spin-coated on the mold, to a thickness of about 100 $\mu$m, and was then cured by baking for 30 minutes at 100$^{\circ}$ C. Then, two PZT actuators (Digikey part number 668-1000-ND) were placed on the pre-defined sorter location on the mold. Additional PDMS was then poured on the mold, step 9, to a total thickness of about 5 mm and cured for 2 hours at 80$^{\circ}$ C. The hardened elastomer was then peeled off of the mold, and ports 1 mm in diameter were punched in the PDMS for fluid access, and the PDMS then cleaned in isopropanol. Following this, the PDMS surface treated for 10 minutes in a UV-ozone discharge immediately prior to bonding to a clean glass substrate. (b) Left: Optical micrographs of the complete device bonded on the glass substrate. Right: shows a magnified view of the region including the electrical sense electrode and the micro-sorter.
\label{fig:6}}
\end{figure*}

The micromold used to define the microfluidic sorter poly(dimethylsiloxane) (PDMS) structure was fabricated using two thicknesses of the photodefinable epoxy SU-8 (MicroChem Corp.), patterned using optical lithography; see Fig. \ref{fig:6}. The substrate consisted of a bare 100 mm silicon wafer onto which we defined gold alignment marks using a photoresist liftoff process. The thin regions of the micromold (2 $\mu$m thick), which includes the fluidic resistor (FR) and microconstriction (MR), were patterned first. The fluidic resistor was defined using 17 parallel ribs of SU-8, each rib 50 $\mu$m long and 3 $\mu$m wide. The microconstriction was a rib of SU-8 2 $\mu$m long and 2 $\mu$m wide. The remaining, thicker portions of the microfluidic structure were defined following this, using thick (15 $\mu$m) SU-8. This mold was used to cast PDMS, also shown in Fig. \ref{fig:6}, and could be reused to fabricate many PDMS elements. Once cast, the PDMS was bonded to a clean glass slide that, for the size-based micro-sorter, included a sensing electrode (10 nm Ti, 50 nm Au), fabricated using a conventional photoresist liftoff process.

\subsection*{Steady-state device operation}
The microfluidic channels were completely filled with either filtered saline or deionized water, and fluid flow was
established by regulating the air pressure at each of the device ports. Usually, IN, FF1 and FF2 ports (refer to Figure \ref{fig:2}(a)) were set to very similar values to create a pressure head on the analyte input side. The beads or cells were introduced into the IN port on the device, FF1 and FF2 flow-focussing ports were used to hydrodynamically focus the input beads towards the OUT B port.
On the other side of the microsorter, OUT A, B, C ports were set at lower pressure than IN port, yielding the bead velocities of 1
mm/s.

\subsection*{Actuator tune-up}
Using the microactuator to successfully drive particles to the desired output port depends on transducer pulse amplitude, pulse timing as well as on the height of the particle in the channel. The time delay from detection to the actuation point is inversely proportional to velocity, so finding a pulse delay that works for the range of expected velocities is clearly necessary. The height of a particle in the channel determines its velocity, due to the Poiseuille distribution of flow; this also controls the effectiveness of the sorting flow pulse, which has the same Poiseuille flow distribution, so that e.g. a slow particle will feel a smaller sorting force than a fast particle. As slower particles are in the sorting region proportionally longer than fast particles, the integrated impulse can be made almost independent of velocity.  In order to achieve the highest fidelity actuation, we tuned each transducer's actuation voltage separately, and ensured that nearly the same end result could be achieved independent of the particle's measured horizontal velocity. Figure \ref{fig:2}(d) shows part of the tune-up for the push mode actuation, in which the voltage applied to one actuator acting by itself was adjusted to give the same displacement as for the other actuator acting by itself (red and green data points). This was verified by applying the same polarity pulse to each actuator, giving a net null displacement (black data points). For example, for the device used here, we found the upper actuator (A1) would balance the lower actuator (A2) if we used pulses of amplitude $+2.5$ V and $+4.5$ V respectively, the difference in voltage reflecting a difference in the transducers or in how they were mounted. A similar tune-up procedure was used for the pull mode actuation.

\subsection*{Operation of electrical size-based sorter}
Constant and opposite voltages $V_{a}$ and $V_{b}$ (typically a few volts in amplitude) were applied to bias electrodes A and B respectively (Fig. 1a). The bias voltages were adjusted in such a way as to keep the average sensor output voltage $V_{\rm out}$ near zero, to avoid electrolytic corrosion of the sensing electrode. The instantaneous voltage $V_{\rm out}$ was monitored and compared to a set threshold voltage, with voltages beyond the threshold triggering the sorter actuation. The trigger (Stanford Research DG535) would in turn generate a square pulse of 10 ms duration and $V_{p-p}$ = 1 V amplitude, which was amplified by a dual-output amplifier (Krohn-Hite 3602M) to generate positive and negative polarity pulses which were applied to the two PZT transducers. The voltages used to drive the PZTs were tuned up as described above.

\subsection*{Operation of fluorescence-based sorter}
For fluorescence-based sorting, which was used designed to sort both fluorescent beads as well as biological cells with dimensions of order 10~$\mu$m, a much larger microfluidic design was used, where in Fig. \ref{fig:2}(b) and (c) the layout was scaled up uniformly by a factor of 10 from the microparticle sorter. In this implementation, the particles to be sorted were illuminated with a mercury lamp through a DAPI filter cube, and the return fluorescent signal captured using a 60$\times$ dry objective before illuminating a charge-coupled device (CCD) camera (QICAM 12-bit, 100 fps). The pixels in the CCD lying inside a software-defined rectangular ``region of interest'' (ROI), shown in Fig. \ref{fig:2}(c), were digitally analyzed and summed in real time to evaluate the ROI total intensity. When a bead passes through the ROI, the summed intensity increases sharply, and this intensity is used to generate a software-programmable delayed trigger pulse on a function generator (Stanford Research Systems DS345, square pulses of 10 ms duration and $V_\textrm{p-p} = 1$ V amplitude). This pulse was amplified by a dual-output amplifier (Krohn-Hite 3602M) to generate positive and negative polarity pulses with amplitudes as per the tune-up procedure described above. The opposite polarity pulses were applied to the two transducers, generating the desired push-pull drag flow in the channel connecting the two transducers. In the absence of an actuation signal, particles in the flow stream would flow into the waste port B in Fig. \ref{fig:2}.

\begin{figure}
\centering
\includegraphics[width=0.35\textwidth]{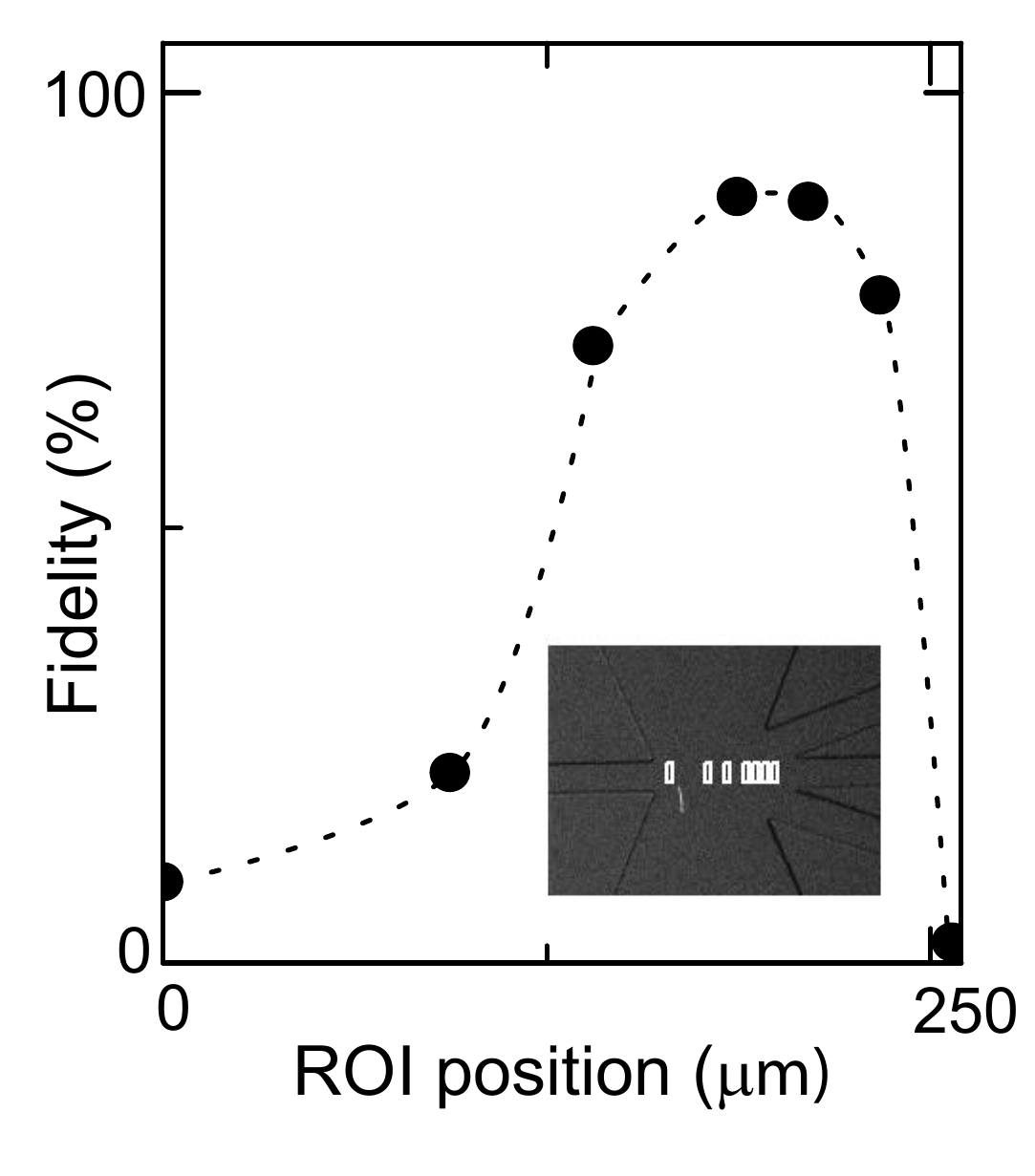}
\caption{\textbf{Fluorescence sorting tuneup:} Fidelity of sorting (percent desired outcome) as a function of lateral (horizontal) position of ROI in camera field of view. Inset shows different positions for the ROI in relation to the microfluidic circuit; the horizontal axis origin is at the leftmost ROI in the inset. \label{fig:5}}
\end{figure}

The fidelity of this sorter, defined as the fraction of fluorescent beads that went to the desired output port, depends strongly on the timing between fluorescence detection and the actuation flow pulse, which required careful optimization due to the larger microfluidic dimensions as well as the use of software analysis and its concomitant delays. We controlled the timing in part by varying the position of the ROI in the camera field-of-view, while monitoring the sorter output, as shown in Fig. \ref{fig:5}. The inset to that figure shows how we varied the lateral position of the ROI in the camera field of view. Using an optimized location for the ROI, we were able in this particular tune-up experiment to achieve a sorting fidelity of 87\%, limited by non-ideal timing control and some variability in the drag force when actuating, due to variations in the height of the beads in the channel. Applying similar optimization procedures to these other parameters, we were able to bring the maximum sorting fidelity up to about 94\%.

\subsection*{Macrophage cell preparation}
J774 murine macrophage (EACC, Salisbury, UK) were stained with Hoeschst 33342 dye, with excitation at 352 nm and emission at 461 nm. Cells were diluted as needed (concentration $~2 \times 10^{5}$ cells/ml) into $1 \times$ PBS containing $2\%$ w/v bovine serum albumin. Cells were shifted through a $75~\mu$m cell strainer and iodixanol density gradient medium was added to a final concentration of $8\%$ w/v. The cell mixture was passed through the microfluidic device while operating the camera at 100 frames per second.

\subsection*{Polystyrene nanoparticle preparation}
Fluorescent polystyrene beads were obtained from Polysciences Inc. (0.5 $\mu$m, catalogue no. 15700-10; 0.75 $\mu$m, catalogue no. 07766-10; and 1 $\mu$m, catalogue no. 15702-10). Number densities in stock solution were calculated from the manufacturer's specifications, and the particles were diluted as needed into 1$\times$ PBS with the addition of $1\%$ Tween 20 (Sigma-Aldrich).

\section*{Acknowledgements}
The authors would like to acknowledge the Keck Foundation for financial support, and the NSF NNIN for support of the UC Santa Barbara Nanofabrication facility. We are grateful to Nelly Traitcheva for mammalian cell culture and cell preparations.

\section*{Author information}
Corresponding author: Email anc@uchicago.edu.

The authors declare they have no competing financial interests.

\end{document}